\documentclass[aps,10pt,prb,onecolumn,final]{revtex4-1}

\usepackage[latin1]{inputenc}
\usepackage{amsmath}
\usepackage{amsfonts}
\usepackage{amssymb}
\usepackage{graphicx}
\usepackage{subfigure}
\usepackage{color}

\begin{document}
\title{Effects of the mean free path and relaxation in a model for the aggregation of particles in superfluid media}

\author{S. G. Alves$^1$}

\author{A. F. Vilesov$^2$}

\author{S. C. Ferreira, Jr.$^1$}
\email{silviojr@ufv.br}

\affiliation{$^1$Departamento de Física - Universidade Federal de Viçosa, 36571-000, Viçosa, Minas Gerais, Brazil \\ $^2$Department of Chemistry, University of Southern California, Los Angeles, California 90089
}

\begin{abstract}
In this paper,  we study a two-dimensional model for the growth of a molecular clusters in superfluid helium at low temperature. In the model, particles of diameter $a$ follow  random ballistic moves of length  $\delta = a - 256a$. Upon attachment on the cluster surface, particles allow one-step relaxation to the nearest two-fold coordinated site. Average coordination numbers of particles show that in the presence of relaxation the screening for incoming particles modifies the microscopic structure of the cluster. These results are in qualitative agreement with experimental aggregation of methane in He droplets, in which large abundance of fully coordinated sites is observed. The average coordination number
increases with  $\delta$, showing that screening is relevant when relaxation is present. 
As the cluster size increases, the corresponding structure clusters changes from a compact to a fractal, typical of ballistic and diffusion limited models, respectively. A scaling ansatz describing the crossover between the two regimes is proposed.

\end{abstract}

\maketitle

\section{Introduction}
\label{intro}
Experiments with atoms and molecules embedded in He droplets of up to $10^8$ He atoms have recently became available. \cite{Toennies} Single molecules were found to rotate freely in He droplets and are unique probes of superfluid helium on a microscopic scale.\cite{Hartmann,Grebenev} He droplets have extensively been used as hosts for the growth and spectroscopic study of small atomic and molecular clusters.\cite{Toennies} It was obtained that experiments in He droplets often supersede other known techniques, such as free nozzle beam expansion, matrix isolation, and cryogenic cooling cells. In particular, He droplets offer straightforward control over the cluster size. One of the most spectacular observations was that of the self assembly of chain clusters of polar linear molecules \cite{Nauta1} and unusual ring clusters,\cite{Nauta2} which demonstrated the deep role of aggregation kinetics on the resulting cluster structure. These observations spawned expectations that He droplets could be used as a host in the formation of unique mesoscopic clusters consisting of several hundred and larger number of particles, which are interesting objects on their own and may be of practical use.  Using of He droplet technique for assembly and investigation of large clusters is emerging.  Formation of metal clusters having up to several thousand atoms in He droplets have been documented.\cite{Bartelt,Doppner}  Recent infrared spectroscopic studies of ammonia \cite{Slipchenko1} and methane \cite{Slipchenko2} clusters in He droplets consisting of up to $10^4$ molecules, both suggested formation of compact clusters. These observations are surprising taking into account the very low temperature of 0.38 K in the droplets, which we thought would favor formation of very porous clusters, as previously reported in the bulk liquid He.\cite{Gordon} 

Previous studies identified a few general models of cluster growth \cite{Meakin} such as ballistic aggregation (BA) \cite{Ferreira2005, Liang} and diffusion limited aggregation (DLA),\cite{Meakin} which give rise to densely branched and fractal clusters, respectively.  However, to our knowledge those structures have never been observed in free atomic or molecular clusters, which are usually obtained at rather high temperature, and the structures - even the number of particles in the clusters - are defined by thermodynamics,\cite{Baletto} rather than aggregation kinetics. Helium droplets seem to be an ideal medium to study atomic and molecular aggregation, which allows the realization of some important paradigms such as adding one particle at a time and avoiding coalescence of clusters. Molecules in superfluid helium are expected to move ballistically, until they recombine or are scattered on some elementary excitation in the superfluid, which are extremely rare at the temperature of the droplets of just 0.38 K.\cite{Hartmann} Thus, the mean free path of the molecules in He droplets should be very long. However, it remains unclear what are superfluidity influences on the kinetics of the cluster formation inside the droplets. 

In order to obtain more insight into the mechanism of low temperature aggregation in He droplets, we calculated the structure of the two dimensional clusters using aggregation models modified in some important features. In standard models it is usually assumed that incoming particles immobilize immediately upon the impact. However, atoms and molecules may retain the ability to relax even at very low temperature due to quantum mechanical zero point delocalization and tunneling. The relaxation is guided towards some local minimum energy structures, which is defined by the bonding directionality, like in the case of covalent and hydrogen bonding, or by the largest coordination number such as in the case of Van der Waals bonding. In order to study the possible effect of superfluidity on the cluster structure, we allowed particles moving randomly with different mean free path before attachment to the cluster.

The paper layout is as follows. In section \ref{model}, the model and the computation details are described. In section \ref{results}, the simulation results are presented and discussed. Finally, some conclusions are drawn in section \ref{conclusions}.

\section{\label{model} The Model}

Morphology of the clusters obtained via aggregation of identical particles has been extensively studied experimentally \cite{Bodea} and theoretically.\cite{Ferreira2005,Ferreira2004,Alves,Castro,Seager} DLA  is probably the most studied aggregation model.\cite{Meakin} According to the standard DLA model,\cite{Witten} particles are released, one at a time, far from a cluster and follow unbiased Brownian motion until they touch the cluster where they instantly immobilize. DLA model generates structures with complex multiscale and multifractal properties.\cite{Mandelbrot} Several generalizations of the DLA model were proposed,\cite{Meakin} including models in which the particles take long random steps.\cite{Ferreira2004,MeakinPRB84} For example, Meakin \cite{MeakinPRB84} replaced the Brownian motion of the DLA model by L\'evy flights with the step length $l$ distributed according to $p(l)\sim l^{-\alpha}$. The Meakin's model exhibits a continuous transition from BA- to DLA-like clusters as the exponent $\alpha$ is varied from 0 tho infinity. More recently, Ferreira \cite{Ferreira2004} studied an extended on-lattice DLA model, in which the particles execute long flights with fixed length at random directions. This procedure reduces the screening\cite{screening} for incoming particles, which is a central feature of the DLA-like models. DLA-like structures were observed for small flight length while a transition between compact and DLA-like structures was observed for long flights.

The relaxation of particles upon attachment to the cluster was recently incorporated into the DLA calculations.\cite{Seager} The calculations lead fractal clusters with dense ``arms'' characterized by a high coordination numbers of the particles as compared to those in the original DLA. The relaxation revealed to be important in aggregation of nanoemulsion droplets\cite{Wilking} as demonstrated by measured structure factor $S(q)$ peaked at high values of the wavenumber $q$, which is typical for a close packed structure.

In the present work, we report the results of the two-dimensional calculations, which are based on an algorithm which has been described previously.\cite{Ferreira2004} Simulations started with a single particle stuck to the origin. Additional particles with diameter $a$ are released one at a time and execute ballistic steps of size $\delta$ at random directions. The particles were launched from a circle of radius $R_l$ which is taken to be $R_l=R_{max}+5\delta$, where $R_{max}$ is the largest distance from the origin among all particles in the cluster. When the walker touches any particle of the cluster its trajectory is terminated and the particle relaxes to the nearest position having two or more bonds. However, if the distance between the particle and the cluster becomes larger than $R_k=100\delta R_{max}/a$, the particle is discarded and a new one is released from the launching circle. Since the asymptotic behavior is typically reached in large clusters of about $10^6-10^7$ particles, an efficient algorithm for off-lattice simulation is required.

We adopted the standard DLA off-lattice simulation strategies.\cite{AlvesBJP} First, we allow the particles outside a circle of radius $R_l+5\delta$ taking long random steps of the length $r_{ext}$ provided that the particle remains outside of the launching circle. We adopted $r_{ext}=\max[r-(R_l+5\delta),\delta]$, where $r$ is the distance of the particle from the origin. Also, the movement in large empty areas in the inner region which surrounds the cluster ($r<R_{max}$) are very computer time consuming, especially for large aggregates. Thus, we used a strategy that allows the particles inside the launching circle to take long steps of length $r_{int}$ if they do not cross any part of the aggregate. 
Finally, the determination of the contacts between the wandering particles and the cluster in the off-lattice simulations is another time consuming task which was performed following algorithms described elsewhere.\cite{AlvesBJP} 

\section{\label{results}Results and discussion}

In contrast to the extensive studies concerning  the mobility of electrons and He$^+$ ions in liquid $^4$He~\cite{Reif,Meyer} and $^3$He,\cite{Sluckin,Roach} the motion of neutral atoms and molecules in liquid helium has not deserved much attention.  It was obtained that their mobility is determined by scattering on roton and phonon elementary excitations in $^4$He,\cite{Meyer} and by quasi-particles at Fermi level in $^3$He.\cite{Sluckin, Roach}   The mean free path of He$^+$ ions  in superfluid $^4$He at 0.5 K was estimated to be about one micron.\cite{Rayfield} In turn, mean free paths of the order of about or less than the distance between helium atoms were obtained close to or above the superfluid transition temperature (2.17 K) in $^4$He\cite{Reif,Meyer} or in nonsuperfluid $^3$He.\cite{Sluckin,Roach} 
The scattering cross section for He$^{+}$ was estimated to be about $5\times10^{-14}$ cm$^2$, which corresponds to an effective radius of about 12 \AA.\cite{Reif} The large size of the ions is due to the strong polarization interaction of the charge and He atoms, resulting in the formation of the ``snowball''.\cite{Schwarz} In comparison, interaction of neutral species with liquid helium is due to weak van der Waals forces, which primarily affects the distribution of He atoms in the first solvation shell. \cite{Toennies}  Therefore the scattering cross section for atoms and small molecules, such as methane should be of the order or less than about $10^{-14}$ cm$^2$.  Thus the mean free path of molecules in liquid $^4$He at $T < 0.5$ K should be about or greater than $10~\mu$m, which is much larger than the particle diameter. Hence, particles experience long ballistic flight before collision with the next elementary excitation.  In contrast, the motion of atoms and molecules at temperatures close to the superfluid transition, as well as in normal fluid $^4$He and  $^3$He, should be diffusive with characteristic mean free paths of about or less than the particle's diameter. In our simulations, these experimental estimates correspond to the parameter $\delta$ varying approximately from 1 up to $10^3$.

Figure \ref{fig:fig1} shows the clusters grown for some values of the step size $\delta$. The simulations were stopped when the cluster reached the border of a square region having size of $500a\times 500a$. As the parameter $\delta$ increases, a continuous morphological transition from ramified to homogeneous clusters is observed. For large values of $\delta$ the patterns are essentially those observed for the ballistic aggregation model but they become very similar to the DLA as $\delta$ decreases. For comparison, clusters obtained via standard DLA and BA calculations without relaxation are shown in Fig. \ref{fig:fig1}(e) and \ref{fig:fig1}(f), respectively. It is seen that relaxation causes thicker branches of the aggregates which widths seem to be dependent on the $\delta$ value. Inserts to panels in Fig. 1 show some examples of the local structure of the particles in the clusters. It is seen that relaxation favors large regions of close packing with some vacancies, contrasting with ordinary BA and DLA calculations which result in very erratic packing.

Figure \ref{fig:fig2} shows the distribution of the number of the nearest neighbors, $p_z$, in clusters consisting of $10^3$ and $10^5$ particles in panels (a) and (b), respectively. For sake of comparison,  Fig. \ref{fig:fig2}(b) also includes the $p_z$ distributions for the original DLA and BA calculations. It is seen that clusters obtained with relaxation have a large abundance of the fully coordinated ($z=6$) and single vacancy sites ($z=5$). In addition, sites having $z =3$ and 4 correspond to the particles on the outer or inner boundary surfaces. The results clearly differ from the clusters obtained via BA and DLA algorithms both having very similar $p_z$ distributions peaking at $z=2$. This indicates that the screening for incoming particles has no effect on the local arrangement of the particles in these models. Insert to Fig. \ref{fig:fig2}(a) shows the probability of the occurrence of the fully coordinated site as a function of $\delta$. For each cluster size, the results indicate a logarithmic increase of the coordination number for small $\delta$, followed by saturation at large $\delta$. This is consistent with the BA-like scaling regime that is expected to be dominant for $\delta$ larger than the characteristic cluster size. These results are in qualitative agreement with experiments on the aggregation in He droplets, which show large abundance of fully coordinated sites. The overall shape of the distribution is shifted towards large coordination number as $\delta$ increases, showing that screening becomes important when the relaxation mechanism is present. The insert to Fig. \ref{fig:fig2}(a) also shows that the fraction of fully coordinated sites in the interior of the cluster remains nearly constant as more particles are added to the aggregate, in agreement with the experiments in He droplets showing compact clusters.

The asymptotic DLA-like behavior is illustrated in figure 3, where sequential stages of the cluster grown with $\delta = 128a$ are shown. At the early stages of the growth, the clusters closely resemble BA structures, which are characterized by homogeneous distribution of the branches. In larger clusters, the aggregates are dendritic resembling the DLA structures. A crossover from BA, at the early stages, to the DLA structures, at asymptotic stages, is observed at some critical cluster size which depends on the step size $\delta$.

An analysis of the cluster shape may provide some additional insights into the system.\cite{Meakin}  Figs. \ref{fig:fig1} and \ref{fig:fig3} suggest a morphological transition from nonfractal to fractal aggregates.  Therefore, characteristics, such as fractality and/or multifractality,\cite{Meakin} as well as, the scaling exponents associated to the clusters \cite{Barabasi} are important parameters describing overall cluster structure. 

In order to characterize the crossover from the BA- to the DLA-like structures, simulations were done on a square region having size of up to $10^4a\times 10^4a$ with the step length ranging from $\delta=1a$ to $256a$. The growth was stopped when the cluster reached the border of the limiting region. The fractal dimension, $d_f$, can be estimated using the mass-radius method by counting of the number of particles $M$ enclosed by a circle of radius $r$ centered at the origin taking the relation
\begin{equation}
M(r) \sim r^{d_f}.
\end{equation}
If $d_f$ equals to the embedding dimension (of two in this work), the clusters are nonfractal. 

Figure \ref{fig:fig4}(a) shows the mass-radius curves for several step sizes. For $\delta >1$, the curves exhibit tenuous crossovers in the power laws, signaling the transition from the BA- to the DLA-like patterns. In all cases the value of $d_f$ is close to $1.96(2)$ in small clusters and turns to $d_f=1.70(1)$ in larger clusters, which correspond to the fractal dimensions of the BA and the DLA aggregates, respectively.\cite{Tolman} Here and later in the paper, the uncertainties in  the last significant digit are shown in parenthesis. The dependences on $\delta$ of the mass $M^*$ and radius $r^*$ of the crossover cluster are shown in the insert to Fig. \ref{fig:fig4}(a). Scaling law relations, namely, $r^*\sim \delta ^z$ and $M^*\sim\delta^\alpha$ with $z=1.26(3)$ and $\alpha=2.39(2)$, respectively, were found.

A quantitative relation describing the crossover may be  achieved using a scaling theory,\cite{Barabasi} which establishes some a functional form for $M(r,\delta)$ involving power laws. We propose a scaling function for the mass distribution:
\begin{equation}\label{scaling_mass}
M(r,\delta) = \delta^\alpha f\left(\dfrac{r}{\delta^z}\right),
\end{equation}
where $f(x) \sim x^{d_f}$, with $d_f=d_{BA}$ for $x\ll1$ and  $d_f=d_{DLA}$ for $x\gg1$.  Here, $d_{BA}$ and $d_{DLA}$ are the fractal dimensions of the BA and DLA clusters. Eq. (\ref{scaling_mass}) assumes that the crossover radius and mass scale as $r^*\sim\delta^z$  and $M^*\sim \delta^\alpha$, respectively, and that the ratio $r/r^*$ rules the dependence of $M$ with the cluster size. If the scaling hypothesis (\ref{scaling_mass}) is correct, the plots $M/\delta^\alpha$ against $r/\delta^z$ for all $\delta$ values must collapse on a universal curve. 

The fractal dimensions used  to determine the crossover points were $d_{BA}=1.96$ and $d_{DLA}=1.71$.\cite{Alves,Tolman} The scaling hypothesis introduced by equation  (\ref{scaling_mass}) is corroborated in figure \ref{fig:fig4}(b), where all data collapsed on a universal curve are shown. Figure 4(a) shows that the mass of a cluster of size smaller than the crossover is independent of $\delta$ implying the scaling relation $\alpha=z d_{BA}$. Indeed, the calculated values of $\alpha =2.39(2)$ and $z d_{BA}=2.47(8)$ are in good agreement. The mass above of the crossover, $r=3000a$, scales as $M\sim\delta^\zeta$ with $\zeta=0.26(3)$, obtained from the insert to Fig. \ref{fig:fig4}(b). Using the scaling behavior of the function  (\ref{scaling_mass}) and the exponents $d_{DLA}$, $\alpha$ and $z$ from the fits, we obtained $\zeta=\alpha-z d_{DLA}=0.24(8)$ which is in very good agreement with the value $\zeta=0.26(3)$ above. Determining the scaling function and the corresponding power laws exponents, we have demonstrated the scale invariance of the clusters.



The periphery or cluster border is defined as the particles forming an external layer impenetrable for incoming particles. Consequently, the spaces between adjacent particles at the border are smaller than a particle diameter. For the DLA clusters, almost all particles of the cluster belong to the border. Therefore, the number of peripheral particles, $S$, increases linearly with the cluster size. It is known that in the BA clusters $S$ scales as $N^{0.67(2)}$.\cite{periphery} The calculated number of peripheral particles as a function of $N$ for different $\delta$ values is shown in figure \ref{fig:fig6}. It is seen that upon increase of $N$, $S$ exhibits a neat crossover from a scaling regime $S\sim N^{0.69}$ to $S\sim N$. The exponent $\nu_{BA} = 0.69$ is very close to that obtained in the original BA model, which corresponds to a growth faster than that observed in the compact and round  Eden clusters having $\nu = 0.5$.\cite{periphery} Collapses of all curves were obtained using a scaling hypothesis equivalent to Eq. (\ref{scaling_mass}):
\begin{equation}
S(N,\delta) = \delta^{\alpha'}g\left(\frac{N}{\delta^{z'}}\right),
\end{equation} 
where $g(x)\sim x^{\nu}$  with $\nu=\nu_{BA}$ for $x\ll 1$ and $\nu=\nu_{DLA}=1 $ for $x\gg 1$. The scaling exponents used in the collapse were $\alpha'=1.62(4)$ and $z'=2.28(7)$.

\section{\label{conclusions}Conclusions}

In this work, we modeled aggregation of spherical atomic or molecular particles in helium droplets. At temperatures of about 0.38 K the droplets are superfluid,\cite{Toennies} which presumably give rise to a long mean free path of the particles. We studied the dependence of the structure of the two dimensional clusters on the size of the random ballistic steps as well as on the relaxation of the particles in the cluster. The probability distribution of the local coordination number was used to measure the microscopic arrangement of the particles in the clusters. In the presence of relaxation the distribution was found to peak at the maximum coordination number of $ z = 6$ contrasting with the standard BA and DLA models which peak at $z = 2$. Moreover, the fraction of particles with the largest coordination number increases with $\delta$ at small and saturates for large values. 

Upon increase of the cluster size the morphology shows a crossover from nonfractal to fractal structures. The radial scaling and the size of the active growing zone were used to characterize this crossover. Scaling ansatz was proposed and successfully described the crossovers observed in the mass-radius and number of peripheral particles calculations, demonstrated by the collapses on universal curves observed for distinct values of the step length. The crossing point which delimits the fractal and non fractal regimes, increases with the mean free path. These results indicate that the structure of the clusters is a sensitive function of the mean free path of the molecules in superfluid He and that superfluid state of the clusters is essential for obtaining compact clusters as observed experimentally.

In the future, we plan on expanding the calculations to 3D aggregation and modifying the standard DLA algorithm by taking into account the inter-particle interaction as well as exploring different relaxation mechanisms. Comparison of the experimental and theoretical results will give information on the kinetics of the motion of particles in liquid helium droplets, the role of inter-particle interactions as well as on relaxation pathways of the growing cluster at very low temperature.

\begin{acknowledgments}
This work was supported by the Brazilian agencies CNPq (Conselho Nacional de Desenvolvimento Científico e Tecnológico) and FAPEMIG (Fundação de Amparo a Pesquisa do Estado de Minas Gerais). This work was partially supported by NSF grant CHE 0513163.
\end{acknowledgments}

\newpage

\textbf{FIGURE CAPTIONS}

\begin{figure}[hbt]
\caption{\label{fig:fig1}Clusters obtained with relaxation of the particles for different walker's step lengths: (a) $\delta=a$, (b) $\delta=4a$, (c) $\delta=32a$, and (d) $\delta=128a$. Original DLA and BA models are shown in (e) and (f), respectively. The square region has a length $L=500a$. The number of particles in the clusters is indicated in each panel. The insets show zooms of the regions indicated by the arrows.}
\end{figure}

\begin{figure}[hbt]

\caption{\label{fig:fig2} (Color on-line) Distribution function of the coordination number in clusters consisting of $N=10^3$, and $N=10^5$ particles in panels (a) and (b), respectively, obtained at different step sizes $\delta$. The results are averages over 100 simulation runs. The inset in (a) shows the fraction of particles having coordination number $z=6$ as a function of $\delta$ for two cluster sizes, $N=10^3$ (squares) and $N=10^5$ (circles).}
\end{figure}

\begin{figure}[h!]
\caption{\label{fig:fig3} A cluster at different stages of the growth for $\delta =128a$. The number of particles in the clusters are shown in each panel. The  boxes in (a) and (f) have lengths of $500a$ and $10^4a$, respectively.}
\end{figure}

\begin{figure}[h!]
\caption{\label{fig:fig4} (a) Mass versus radius for $\delta=a, 4a, 16a, 64a$ and $256a$ from the bottom to the top. (b) Collapse of the data shown in panel (a). The inset in (a) shows the crossover radius $r^*$, the mass for $r=r^*$, and the power law fits (dashed lines). In (b), the inset shows a typical curve of the mass above the crossover point ($r=3000a$) as a function of $\delta$. The dashed line has a slope $\zeta=0.30$. Averages were done over 100 runs on a square region of size $5000a\times 5000a$ in the main figures. 10 samples in a region of size $10^4a\times 10^4a$ were used in order to obtain the insets.}
\end{figure}

\begin{figure}[h!]
\caption{\label{fig:fig6} Number of peripheral particles as a function of the total number of particles in the aggregate for $\delta=a, 4a, 16a, 64a$ and $256a$ from the bottom to the top.  The top inset shows the crossover total number of particles $N^*$ and the number of peripheral particles $S^*$. The bottom inset shows a typical curve for $S^*$ above the crossover point ($N=2\times 10^6$) as a function of $\delta$. Averages as in Fig. \ref{fig:fig4}. }
\end{figure}

\newpage

\begin{center}
\includegraphics[width=4.25cm,clip=true]{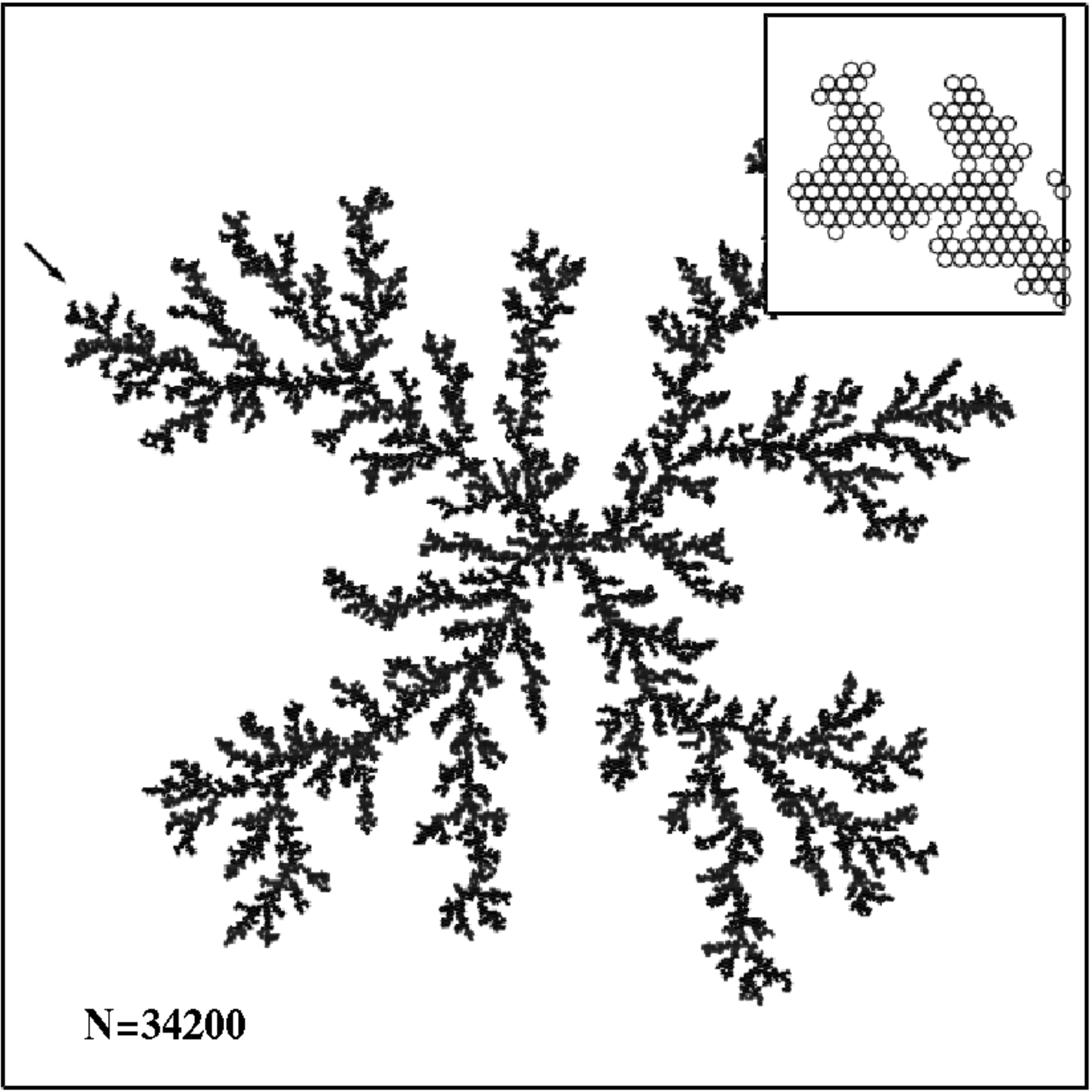}
\includegraphics[width=4.25cm,clip=true]{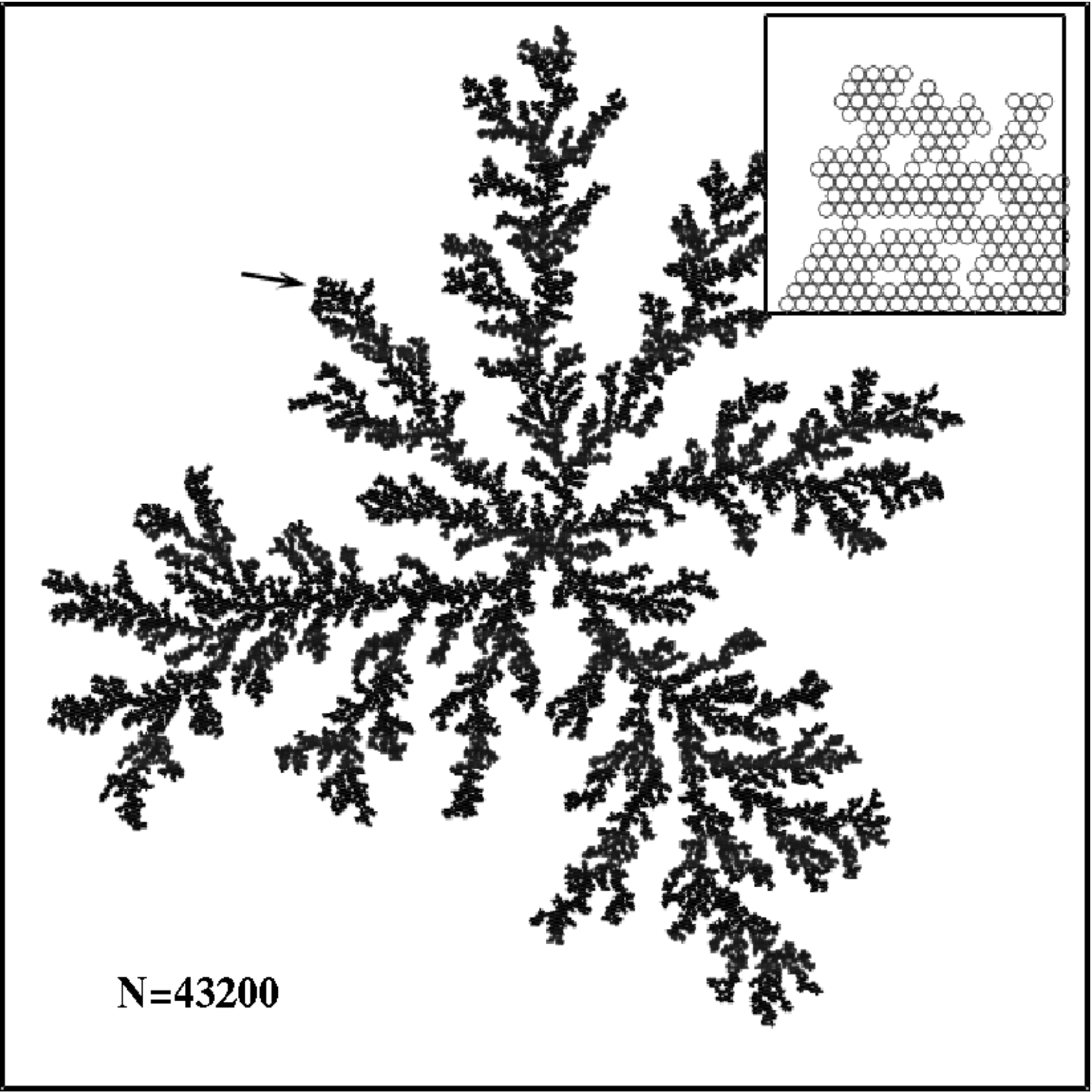}\\
(a)\hspace{3.7cm}(b)\\
\includegraphics[width=4.25cm,clip=true]{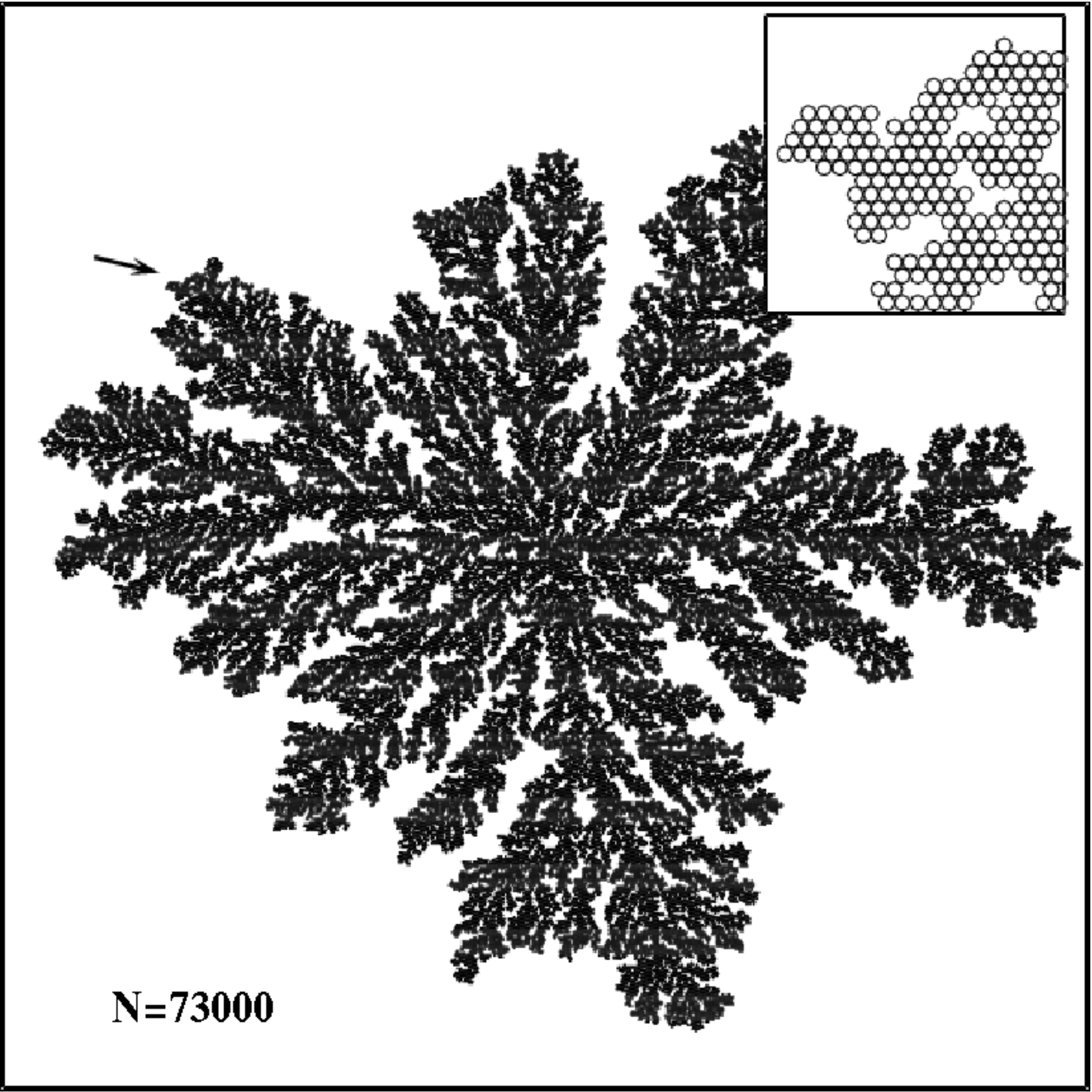}
\includegraphics[width=4.25cm,clip=true]{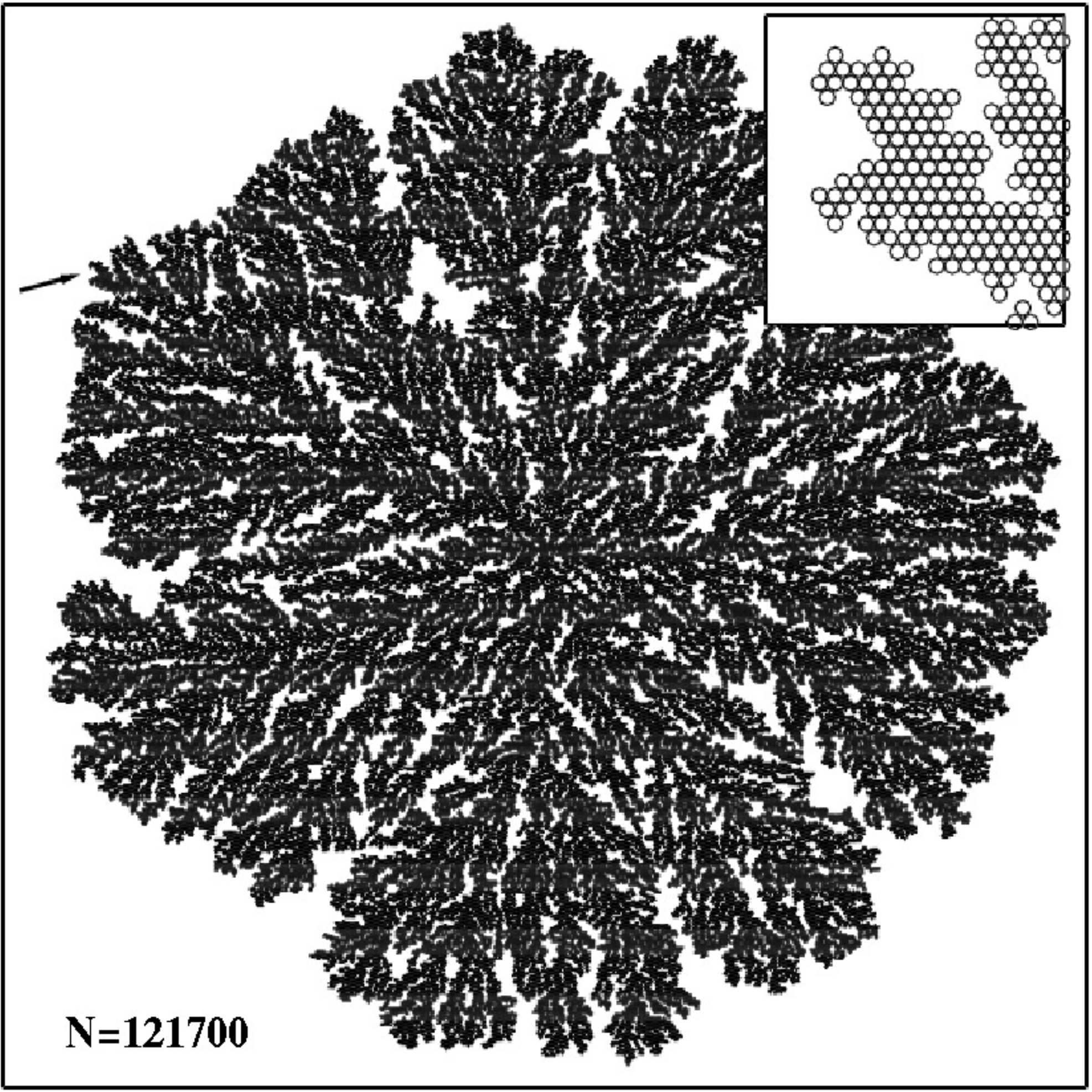}\\
(c)\hspace{3.7cm}(d)\\
\includegraphics[width=4.25cm,clip=true]{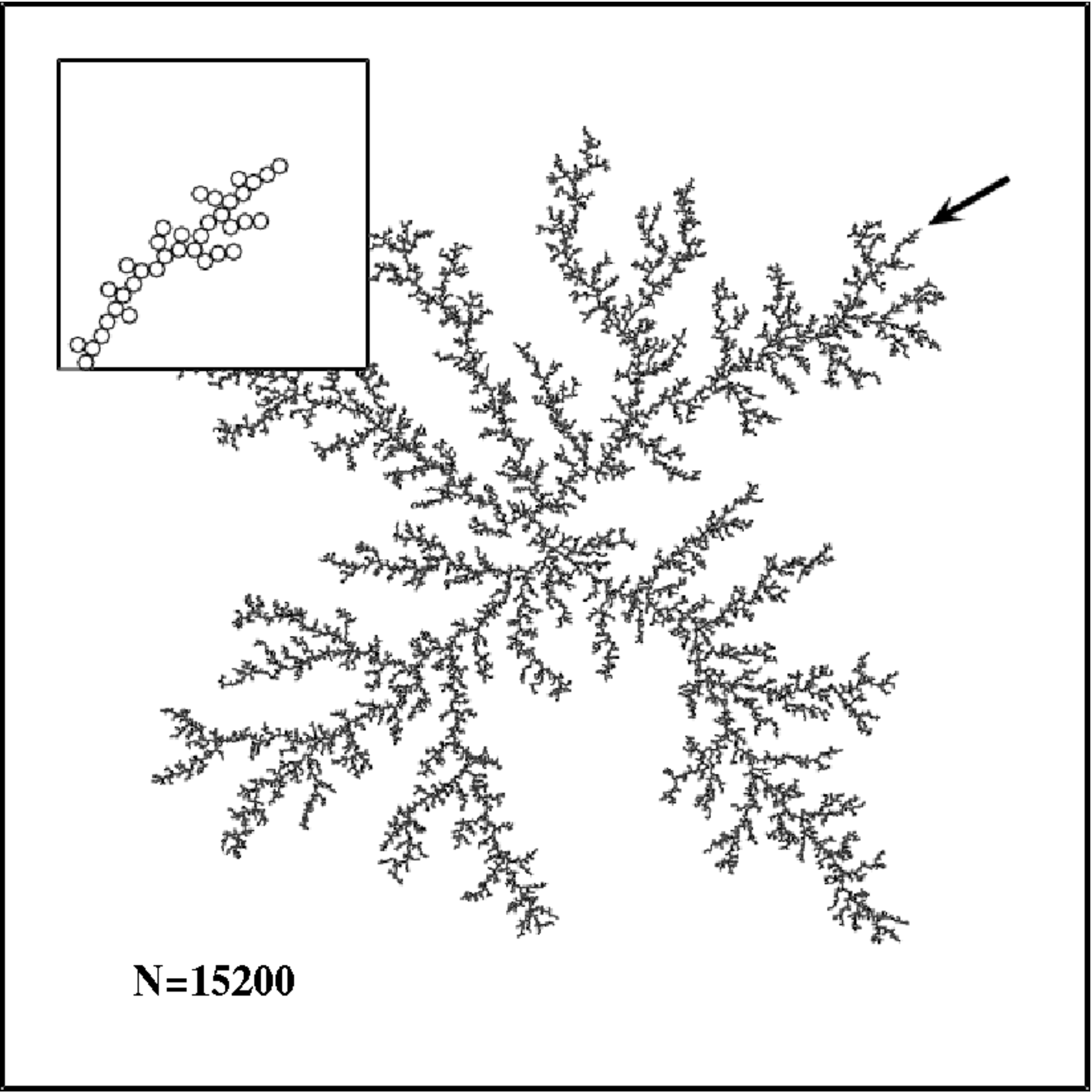}
\includegraphics[width=4.25cm,clip=true]{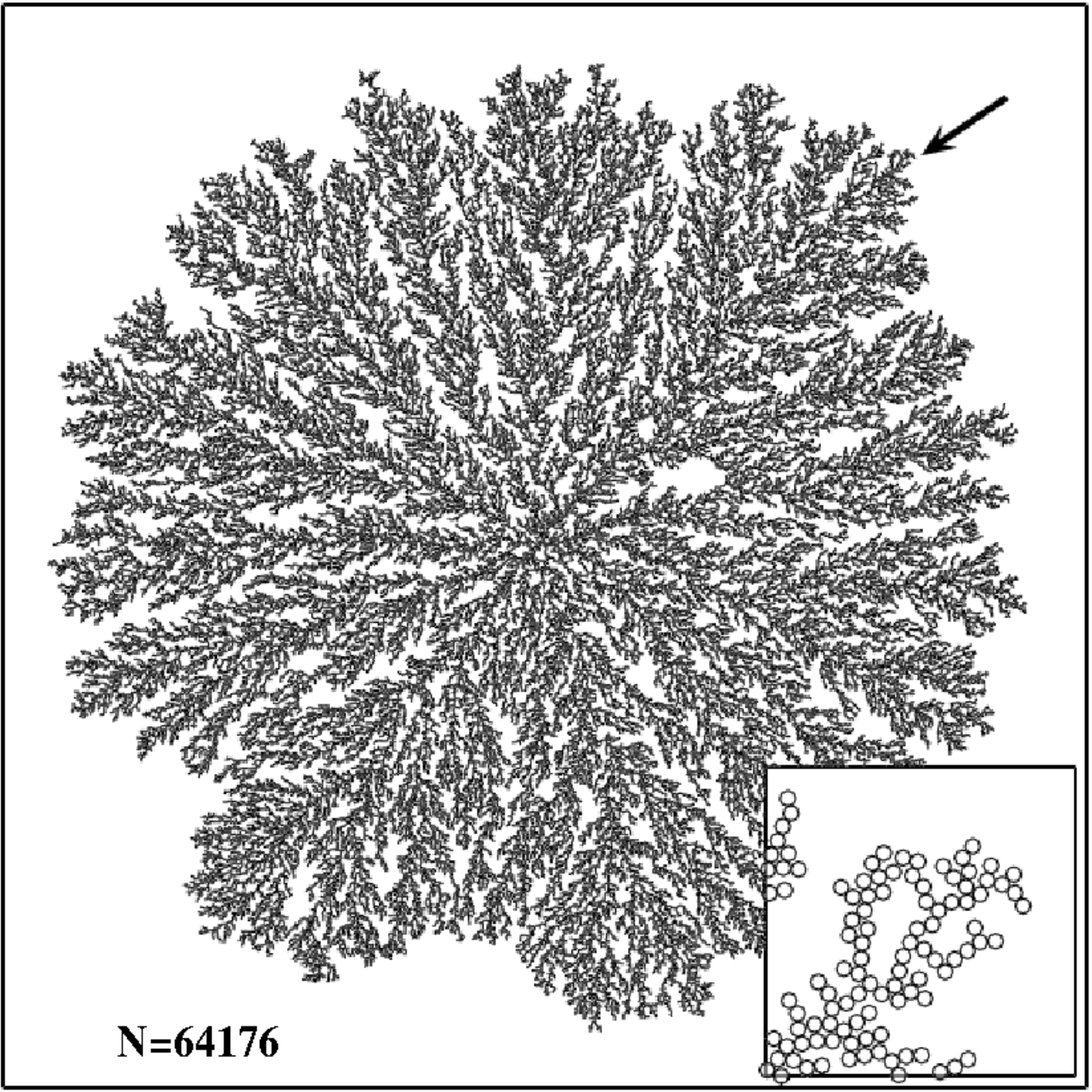}\\
(e)\hspace{3.7cm}(f) \\
FIG. 1
\end{center}

\newpage

\begin{center}
\includegraphics[width=7cm,height=!,clip=true]{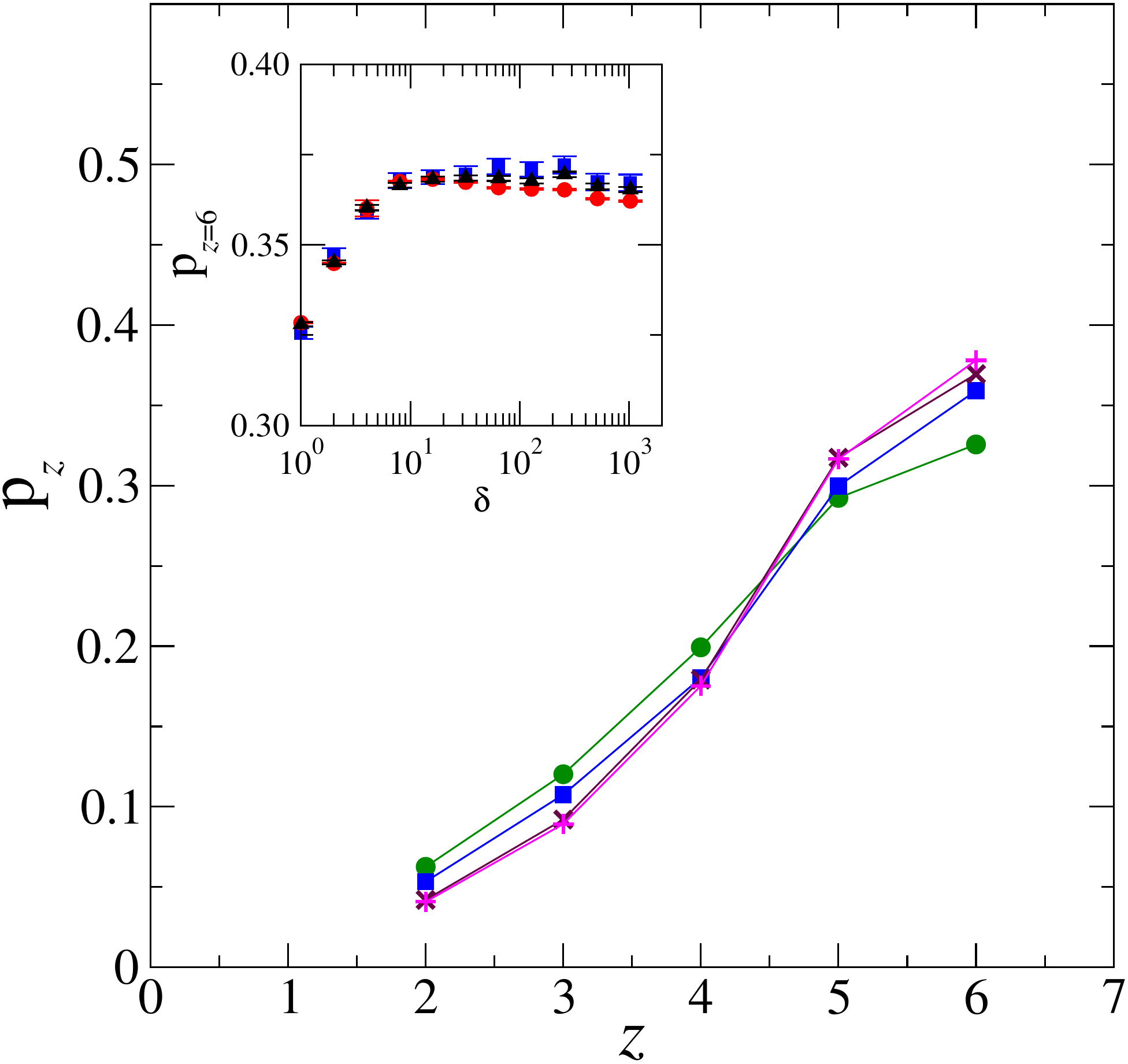} \\ (a) \\
\includegraphics[width=7cm,height=!,clip=true]{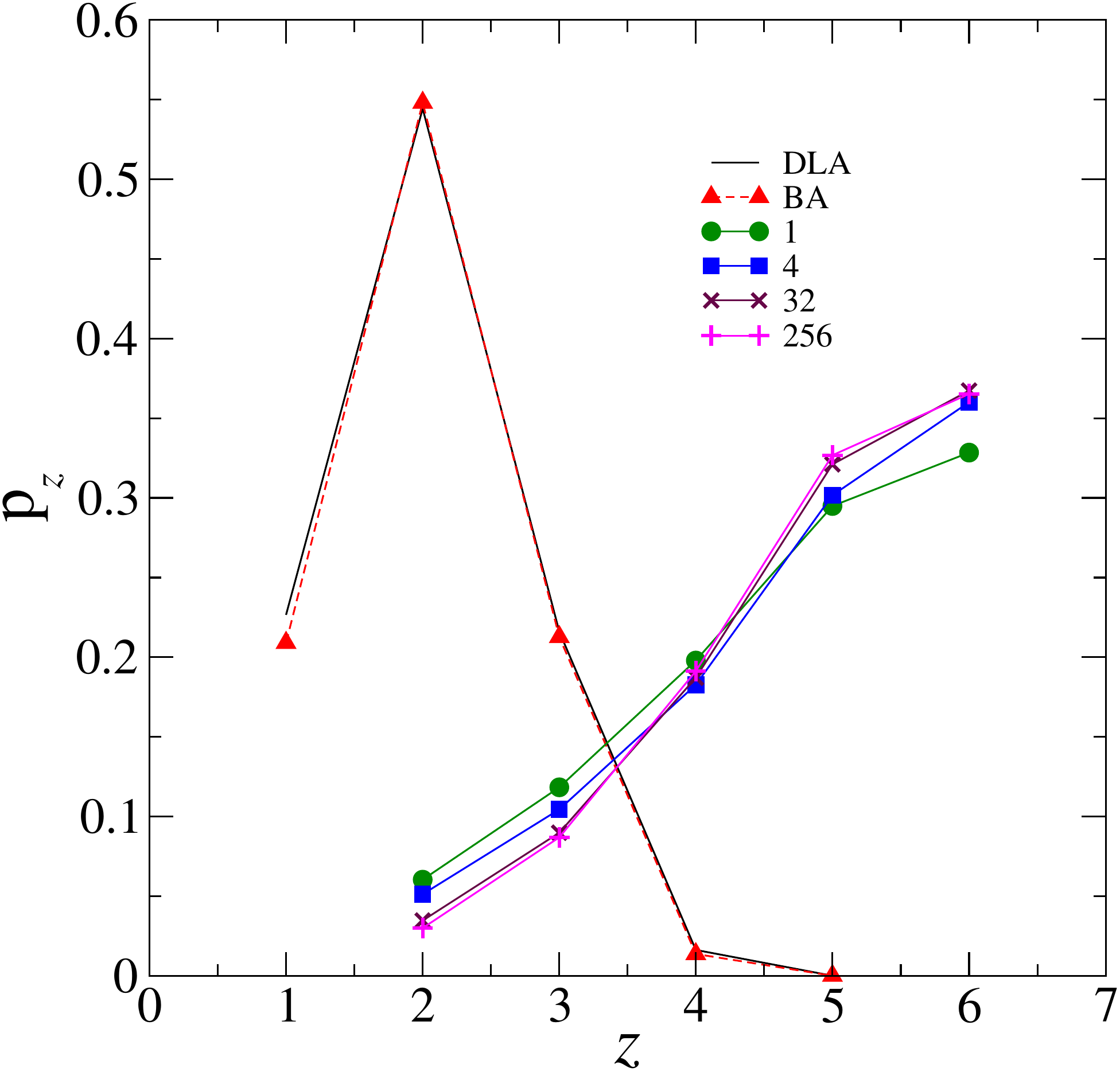}\\ (b) \\ FIG. 2 
\end{center}

\newpage

\begin{center}
\fbox{\includegraphics[width=4.0cm,clip=true]{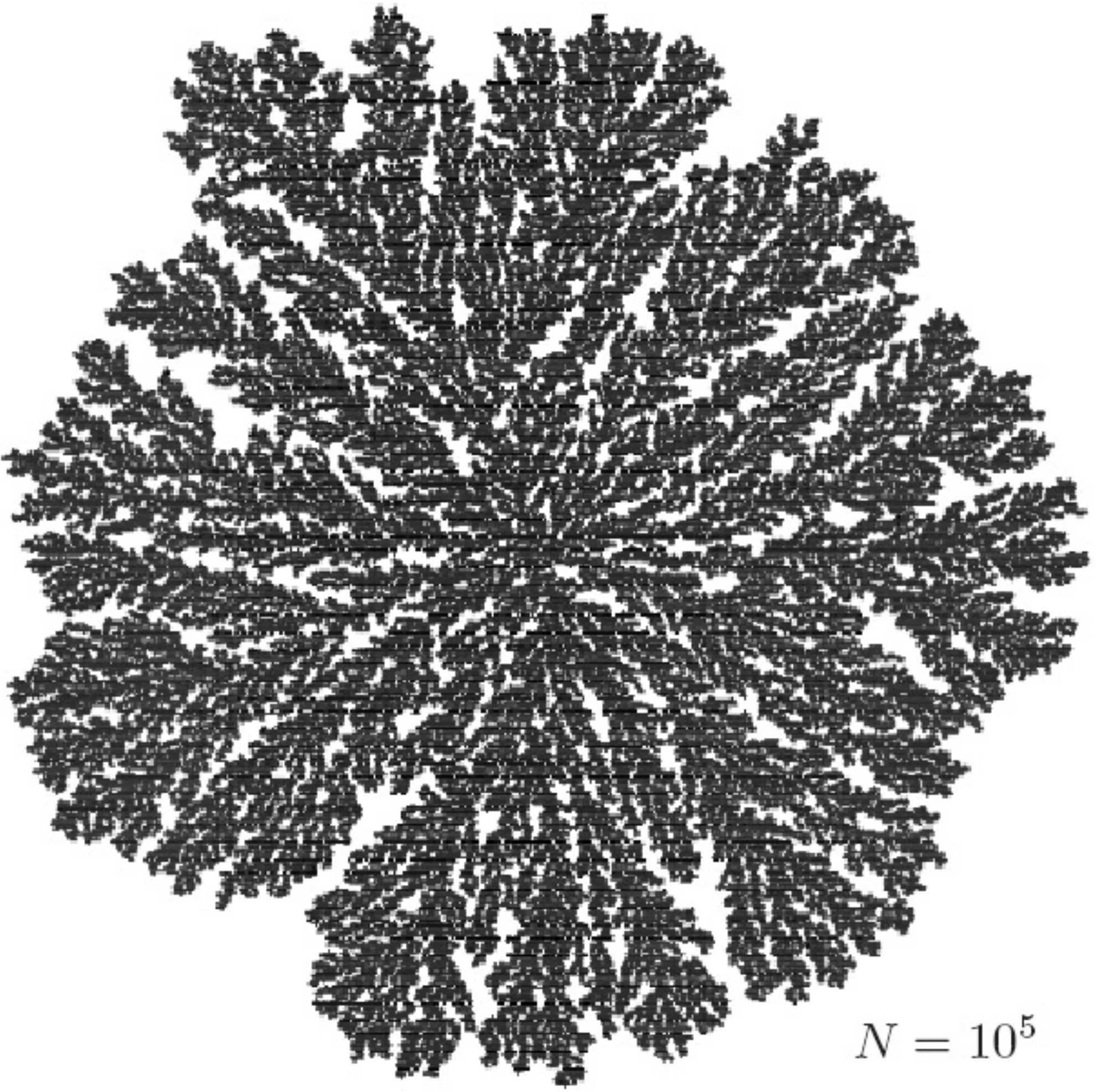}}~
\fbox{\includegraphics[width=4.0cm,clip=true]{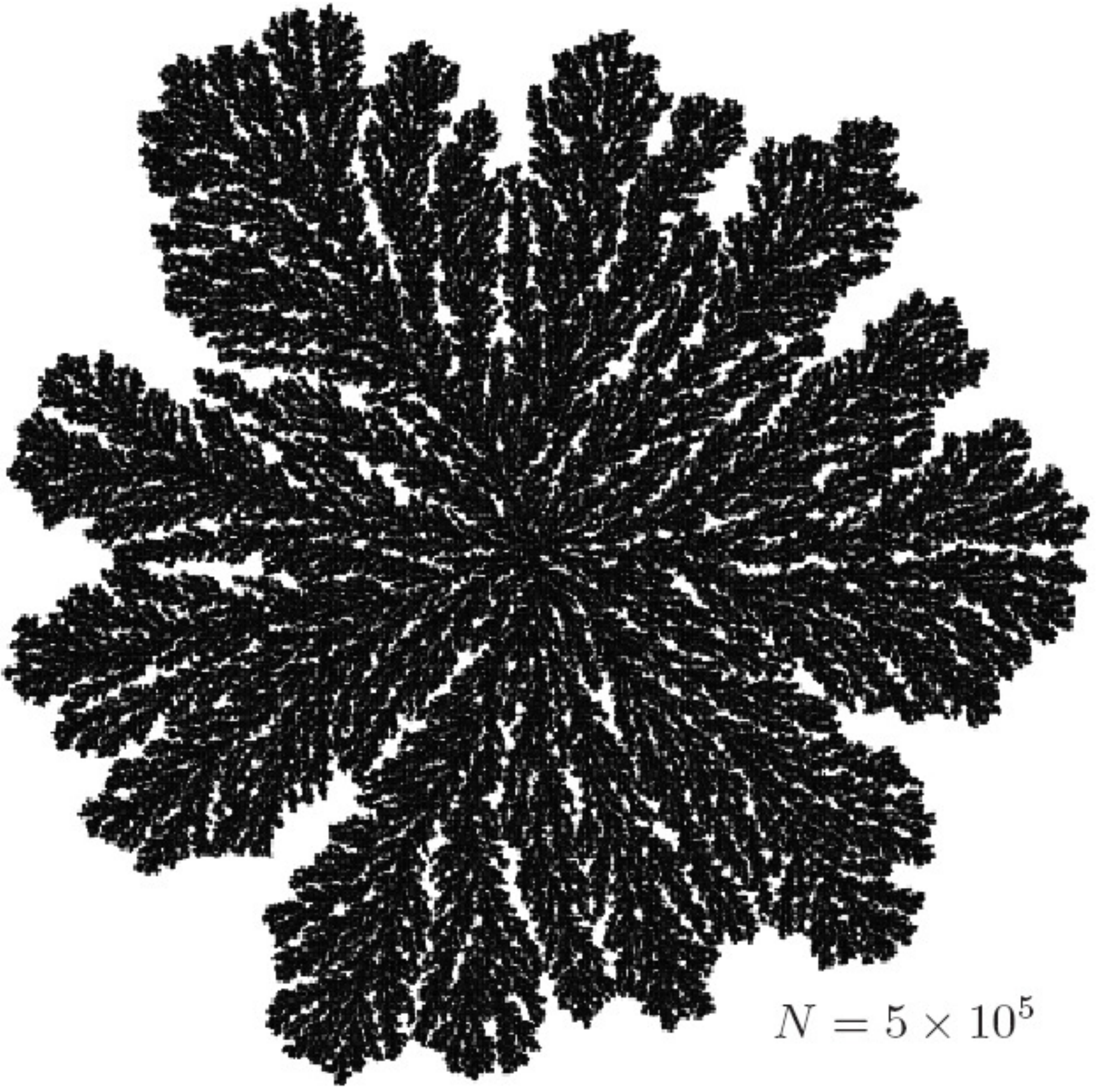}} \\ (a)\hspace{3.5cm}(b) \\
\fbox{\includegraphics[width=4.0cm,clip=true]{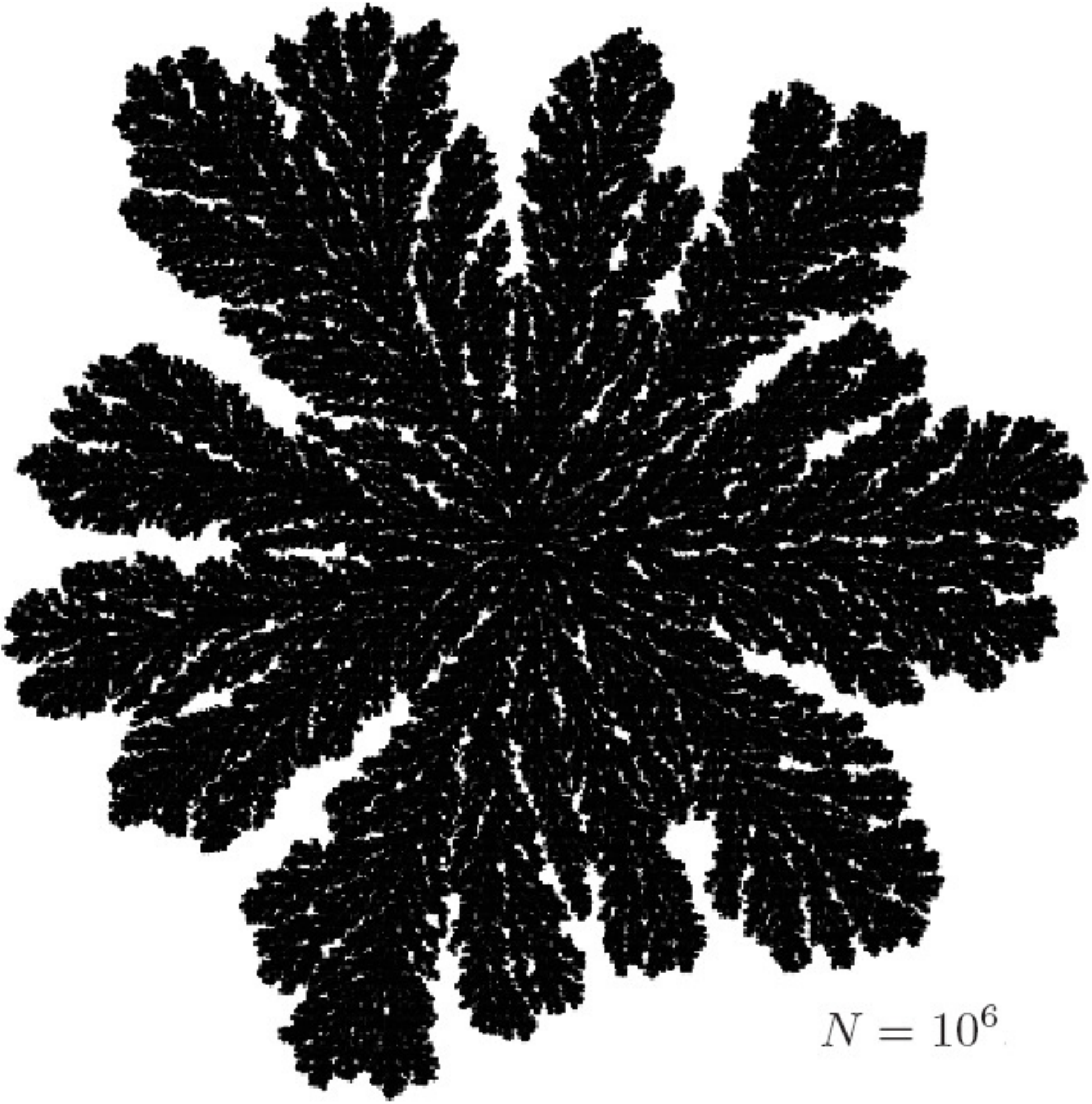}}~
\fbox{\includegraphics[width=4.0cm,clip=true]{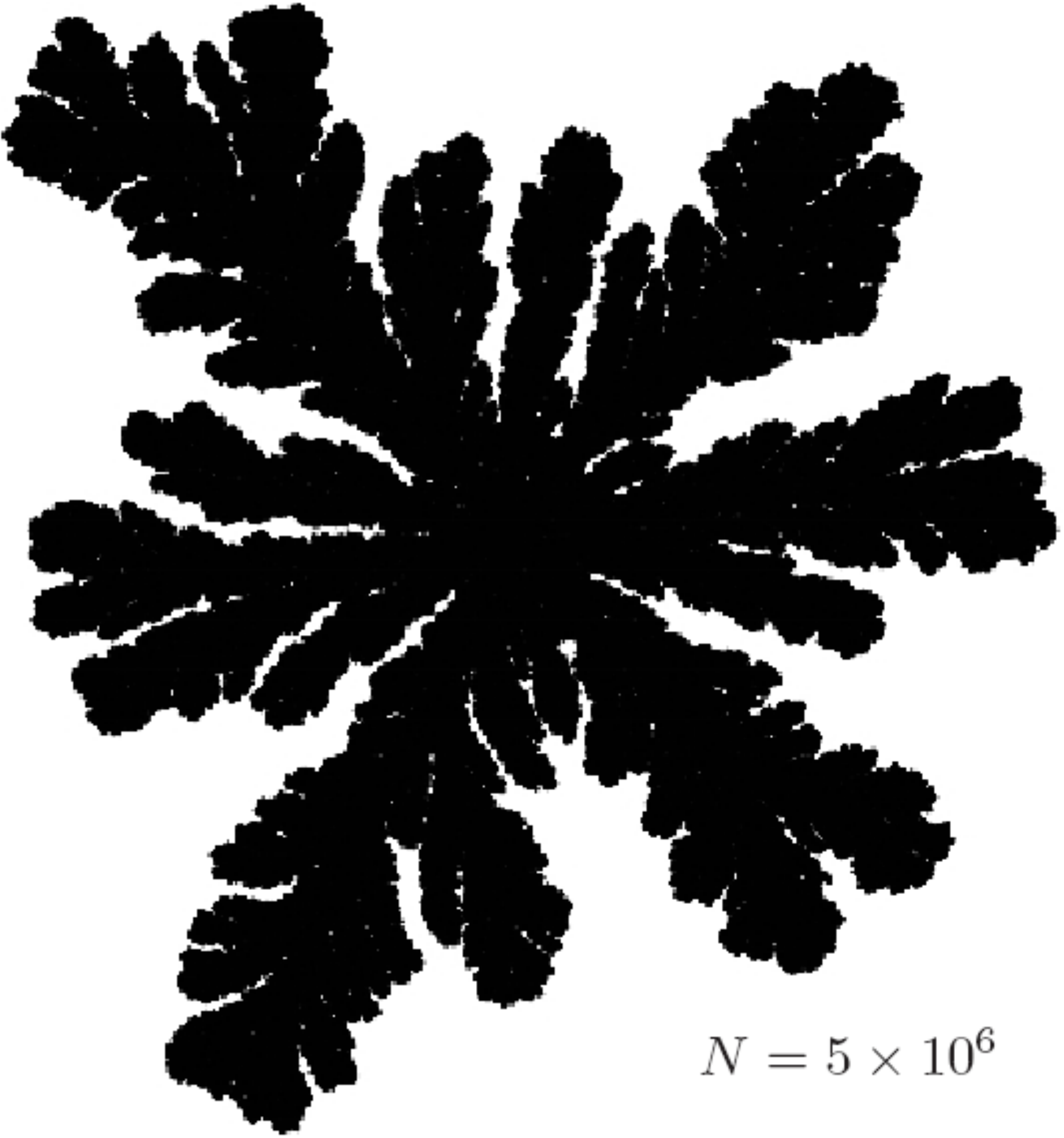}} \\ (c)\hspace{3.5cm}(d) \\
\fbox{\includegraphics[width=4.0cm,clip=true]{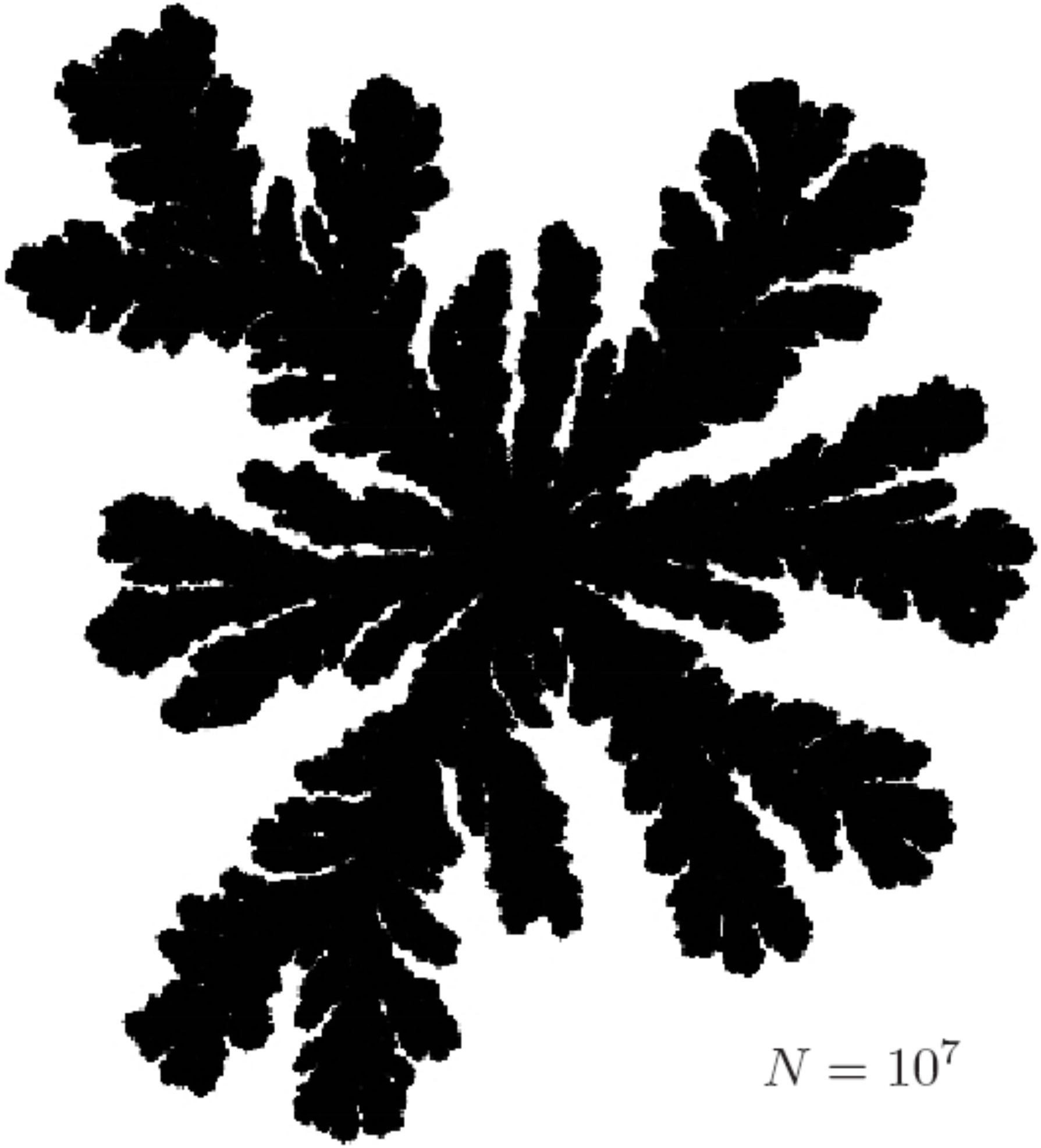}}~
\fbox{\includegraphics[width=4.0cm,clip=true]{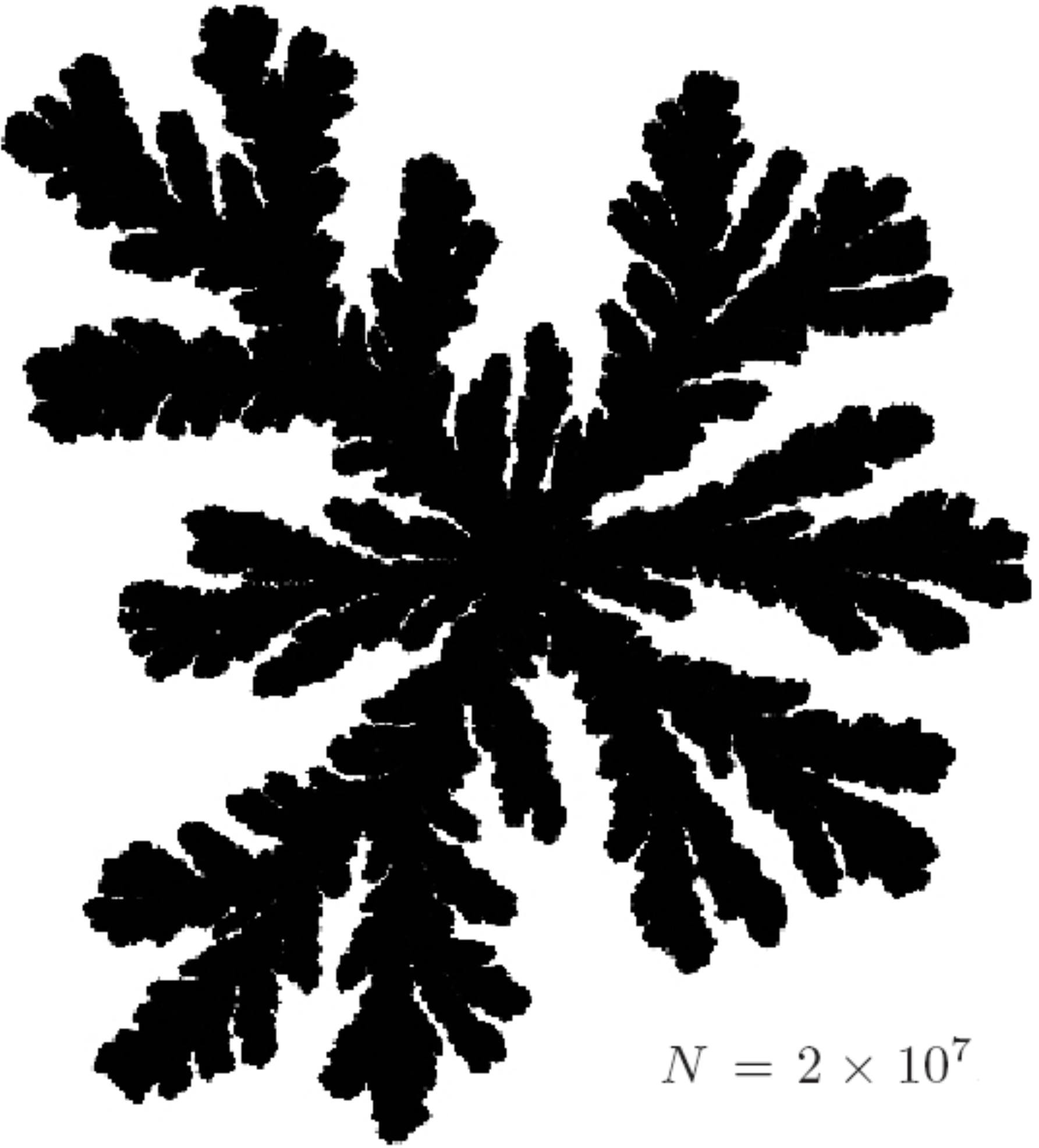}} \\ (e)\hspace{3.5cm}(f) \\FIG. 3 
\end{center}

\newpage

\begin{center}
\includegraphics[width=8cm,height=!,clip=true]{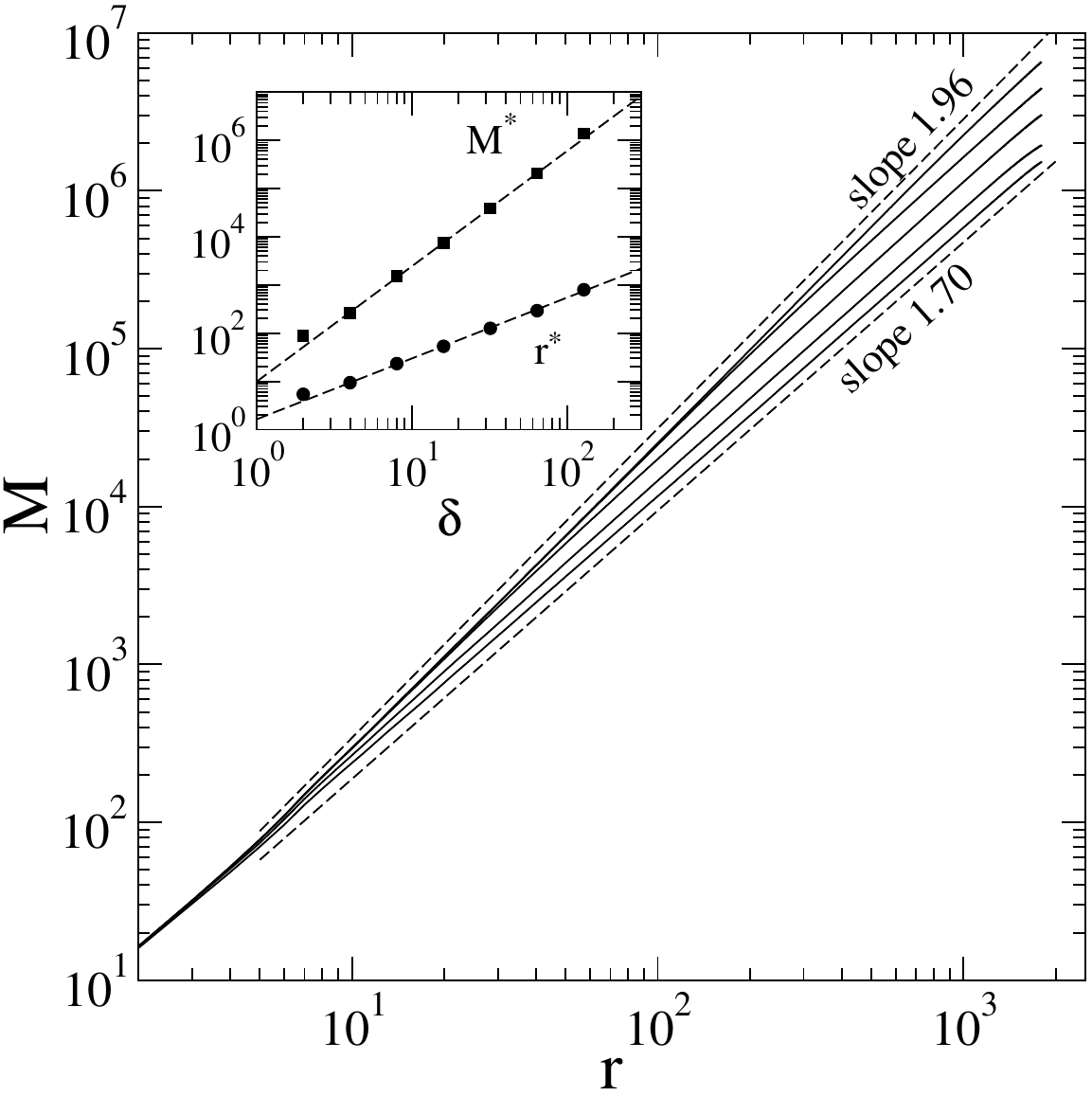} \\ (a) \\
\includegraphics[width=8cm,height=!,clip=true]{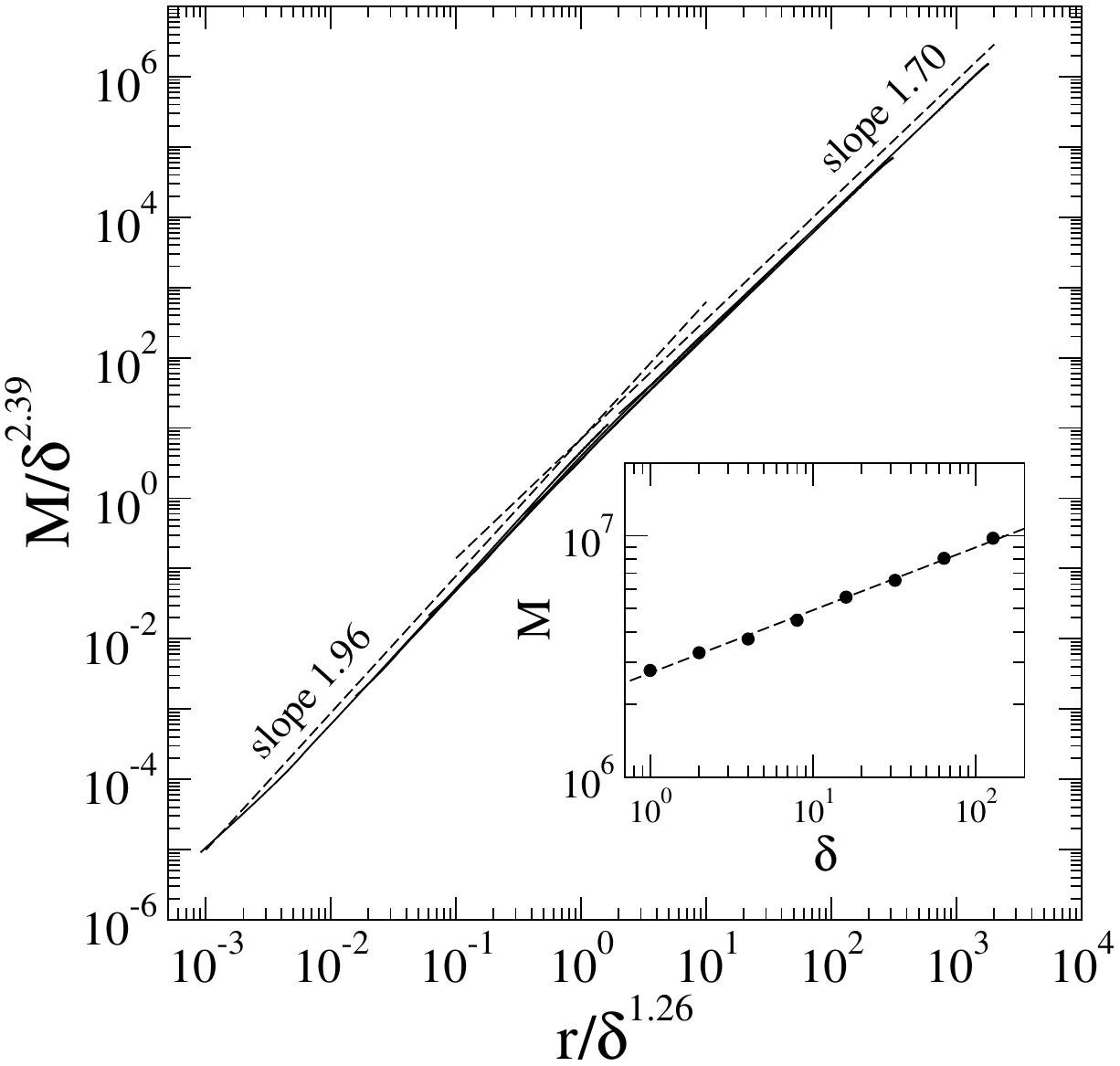}\\ (b) \\ FIG. 4
\end{center}

\newpage
\begin{center}
\includegraphics[width=8cm,height=!,clip=true]{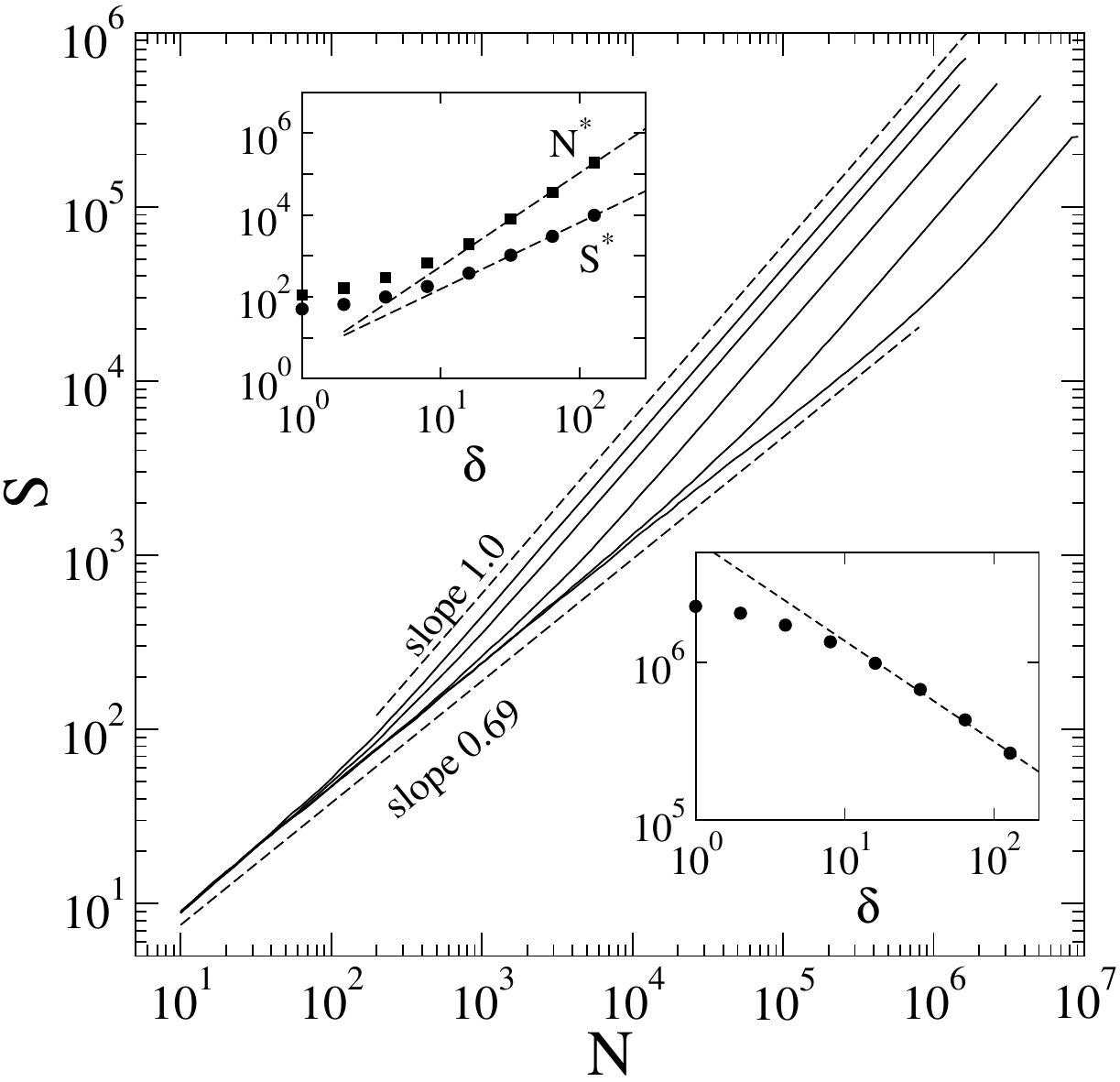} \\ FIG. 5
\end{center}


\begin{thebibliography}{99}
\bibitem{screening} DLA-like models have a tip instability which concentrates the growth to the outermost region of the aggregates and, consequently, resulting a screening for particles incoming to inner regions of the clusters.
\end{thebibliography}
\end{document}